\documentstyle[11pt,epsfig]{article}
\def\edth{\;\raise1.0pt\hbox{$'$}\hskip-6pt\partial\;}
\def\baredth{\;\overline{\raise1.0pt\hbox{$'$}\hskip-6pt
\partial}\;}
\def\gsim{~\rlap{$>$}{\lower 1.0ex\hbox{$\sim$}}}

\def\be{\begin{equation}}
\def\ee{\end{equation}}
\def\ba{\begin{eqnarray}}
\def\ea{\end{eqnarray}}

\newcommand{\omde}{\omega_{de}}
\newcommand{\fr}[2]{\frac{#1}{#2}}

\begin{document}

\title{Growth index with the exact analytic solution of sub-horizon scale linear perturbation for dark energy models with constant equation of state}

\author{Seokcheon Lee$^{\,1,2}$ and Kin-Wang Ng$^{\,1,2,3}$}

\maketitle

$^1${\it Institute of Physics, Academia Sinica,
Taipei, Taiwan 11529, R.O.C.}

$^2${\it Leung Center for Cosmology and Particle Astrophysics, National Taiwan University, Taipei, Taiwan 10617, R.O.C.}

$^3${\it Institute of Astronomy and Astrophysics,
 Academia Sinica, Taipei, Taiwan 11529, R.O.C.}

\begin{abstract}
Three decades ago Heath found the integral form of the exact analytic growing mode solution of the linear density perturbation $\delta$ on sub-horizon scales including the cosmological constant or the curvature term. Recently, we obtained the exact analytic form of this solution in our previous work \cite{SK}. Interestingly, we are able to extend this solution for general dark energy models with the constant equation of state $\omega_{de}$ in a flat universe. This analytic solution provides the accurate and efficient tools for probing the properties of dark energy models such as the behavior of the growth factor and the growth index. We investigate the growth index
and its parameter at any epoch with this exact solution for different dark energy models and find that the growth index is quite model dependent in the redshift space, $0.25 \leq z \leq 1.5$, so observations of the structure growth around this epoch would be very interesting. Also one may be able to rule out some dark energy models by using the analysis from this exact solution. Thus, the analytic solution for the growth factor provides the very useful tools for future observations to constrain the exact values of observational quantities at any epoch related to the growth factor in the dark energy models.

\end{abstract}

The background evolution equations in a flat Friedmann-Robertson-Walker universe ($\rho_m + \rho_{de} = \rho_{cr}$) are \ba H^2 \equiv \Bigl(\fr{\dot{a}}{a}\Bigr)^2 &=& \fr{8\pi G}{3}(\rho_{m} + \rho_{de}) = \fr{8 \pi G}{3} \rho_{cr} \, , \label{H} \\ 2 \fr{\ddot{a}}{a} + \Bigl(\fr{\dot{a}}{a}\Bigr)^2 &=& - 8 \pi G \omega_{de} \rho_{de} \, , \label{dotH} \ea where $\omega_{de}$ is the equation of state (eos) of dark energy, $\rho_{m}$ and $\rho_{de}$ are the energy densities of the matter and the dark energy, respectively. We consider the constant $\omega_{de}$. The sub-horizon scales linear perturbation equation with respect to the scale factor $a$ is well known \cite{Bonnor}, given by \be \fr{d^2 \delta}{da^2} + \Biggl( \fr{d \ln H}{d a} + \fr{3}{a} \Biggr) \fr{d \delta}{d a} - \fr{4 \pi G \rho_{m}}{(aH)^2} \delta = 0 \, . \label{dadelta} \ee We are able to find the exact analytic growing mode solution of $\delta$ for any value of the constant $\omega_{de}$. After replacing new parameters $Y = Q a^{3 \omega_{de}}$ and $Q = \fr{\Omega_{m}^{0}}{\Omega_{de}^{0}}$  in Eq. (\ref{dadelta}), we get \be Y \fr{d^2 \delta}{dY^2} + \Bigl[1 + \fr{1}{6 \omega_{de}} - \fr{1}{2(Y+1)} \Bigr] \fr{d \delta}{dY} - \Bigl[\fr{1}{6 \omega_{de}^2 Y} - \fr{1}{6 \omega_{de}^2 Y(Y+1)} \Bigr] \delta = 0 \, . \label{dhY} \ee A trial solution is $\delta(Y) = Y^{\alpha} B(Y)$ because it is the most general combination of the solution for the above equation (\ref{dhY}). We replace $\delta$ into Eq (\ref{dhY}) to get  \ba && Y (1 + Y) \fr{d^2B}{dY^2} + \Biggl[ \fr{3}{2} - \fr{1}{6 \omega_{de}} + \Bigl( 2 - \fr{1}{6 \omega_{de}} \Bigr) Y \Biggr] \fr{d B}{dY} + \Biggl[ \fr{(3 \omega_{de} +2)(\omega_{de} -1)}{12 \omega_{de}^2} \Biggr] B = 0 \, , \nonumber \\ && {\rm when} \,\,\,\, \alpha = \fr{1}{2} - \fr{1}{6 \omega_{de}} \,\, .  \label{dskY} \ea The above equation becomes the so called ``hypergeometric" equation when we replace $Y = -Y$ with the complete solution \cite{Morse}, \ba B(Y) &=& c_{1} F [\fr{1}{2} - \fr{1}{2\omega_{de}}, \fr{1}{2} + \fr{1}{3 \omega_{de}}, \fr{3}{2} - \fr{1}{6 \omega_{de}}, -Y] \nonumber \\ && \, + \, c_{2} Y^{\fr{1-3\omega_{de}}{6 \omega_{de}}} F[-\fr{1}{3\omega_{de}}, \fr{1}{2 \omega_{de}}, \fr{1}{2} + \fr{1}{6 \omega_{de}}, -Y] \, , \label{B} \ea
where $F$ is the hypergeometric function.
\begin{center}
\begin{figure}
\vspace{1.5cm}
\centerline{
\psfig{file=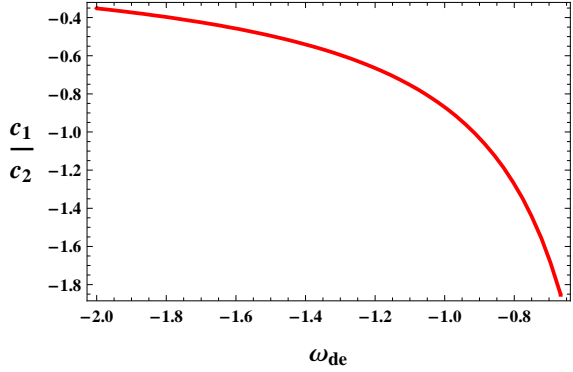, width=7.5cm}}
\vspace{-0.2cm}
\caption{ Behavior of $\fr{c_{1}}{c_{2}}$ as a function of $\omde$ when $\Omega_{m}^{0} = 0.3$ and $a_{i} = 0.1$. } \label{fig1}
\end{figure}
\end{center}
Thus, the exact analytic solution of the sub-horizon scales linear perturbation becomes \ba \delta(Y) &=& c_{1} Y^{\fr{3 \omega_{de} -1}{6 \omega_{de}}} F [\fr{1}{2} - \fr{1}{2\omega_{de}}, \fr{1}{2} + \fr{1}{3 \omega_{de}}, \fr{3}{2} - \fr{1}{6 \omega_{de}}, -Y] \nonumber \\ && \, + \, c_{2} F[-\fr{1}{3\omega_{de}}, \fr{1}{2 \omega_{de}}, \fr{1}{2} + \fr{1}{6 \omega_{de}}, -Y] \, . \label{deltask} \ea This analytic solution does not have any physical meaning before we fix the coefficients $c_{1}$ and $c_{2}$. If we want to have the correct growing mode solution from the above analytic solution, then this solution should follow the behavior of growing mode solution at an early epoch, say $a_{i} \simeq 0.1$. In other words, the coefficients of the general solution should be fixed by using the initial conditions of the growth factor, \be \delta_g(a_i) = a_{i} \hspace{0.2in} {\rm and} \hspace{0.2in} \fr{d \delta_g}{da} \Bigl|_{a_{i}} = 1 \, . \label{ini} \ee For example, $(c_1, c_2) = (-0.716894, 1.07822)$ for $\omega_{de} = -1.2$. We do need these accurate numbers to show the proper growing mode behavior. After we fix the coefficients from the initial conditions, we are able to determine the growth factor $\delta_{g}$ from the general form of solution $\delta$.

\begin{center}
\begin{figure}
\vspace{1.5cm}
\centerline{
\psfig{file=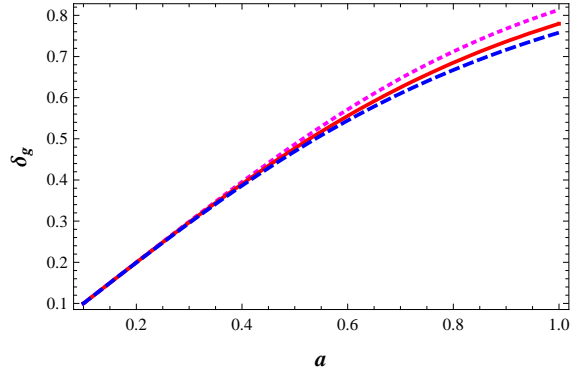, width=7.5cm}}
\vspace{-0.2cm}
\caption{ Evolutions of $\delta_{g}(a)$ when $\Omega_{m}^{0} = 0.3$ for $\omega_{de} = -1.2, -1.0$, and $-0.9$ (from top to bottom). } \label{fig2}
\end{figure}
\end{center}

In the literature, it is used to normalizing the growth mode with respect to its present value. We can obtain $\fr{\delta(Y)}{\delta(Y_0)}$ from Eq. (\ref{deltask}), where $Y_0 (= Q)$ means the value of $Y$ at the present time. Therefore, we only need to know the ratio of $\fr{c_{1}}{c_{2}}$ to determine the growth mode with the initial conditions Eq. (\ref{ini}). From Eqs. (\ref{deltask}) and (\ref{ini}), it is straightforward to show that
\ba && \fr{c_1}{c_2} \Bigl(a_{i}, Q, \omde \Bigr) = 2 a_{i}^{\fr{1-3\omde}{2}} Q^{\fr{1}{6\omde}-\fr{1}{2}} (9\omde -1) \Biggl( -(1+3\omde) \times \nonumber \\ && F \Bigl[-\fr{1}{3\omde},\fr{1}{2\omde},\fr{1}{2}+\fr{1}{6\omde},-a_{i}^{3\omde}Q \Bigr] \nonumber + 3 a_{i}^{3 \omde} Q F \Bigl[1-\fr{1}{3\omde}, 1+\fr{1}{2\omde}, \fr{3}{2}+\fr{1}{6\omde}, -a_{i}^{3\omde} Q \Bigr] \Biggr) \nonumber \\ && \Bigg/ 3(3\omde+1)(\omde-1) \Biggl( a_{i}^{3\omde} Q (3\omde+2) F \Bigl[\fr{3}{2}-\fr{1}{2\omde},\fr{3}{2}+\fr{1}{3\omde},\fr{5}{2}-\fr{1}{6\omde},-a_{i}^{3\omde}Q \Bigr] \nonumber \\ && + (1-9\omde) F \Bigl[-\fr{1}{2\omde}+\fr{1}{2},\fr{1}{2}+\fr{1}{3\omde},\fr{3}{2}-\fr{1}{6\omde},-a_{i}^{3\omde}Q \Bigr] \Biggr) \, . \label{c1c2} \ea
Even though $\fr{c_{1}}{c_{2}}$ is a function of three quantities $a_{i}$, $Q$ ({\it i.e.} $\Omega_{m}^{0}$), and $\omde$, its dependence on $a_{i}$ and $Q$ is quite small. When we vary both $a_{i}$ and $\Omega_{m}^{0}$ from $0.01$ to $0.1$ and from $0.2$ to $0.35$, respectively, the changes on $\fr{c_{1}}{c_{2}}$ are about $10^{-4}$ \% both cases. In Fig. \ref{fig1}, the behavior of the ratio of two coefficients in the growth mode solution is depicted as a function of $\omde$. As $\omde$ decreases, the magnitude of the ratio $\fr{c_{1}}{c_{2}}$ also decreases. This can be better understood with Fig. \ref{fig2}.
In Fig. \ref{fig2}, we show the behaviors of the growth factor $\delta_{g}$ for the different dark energy models ({\it i.e.} for the different values of $\omega_{de}$) when $\Omega_{m}^{0} = 0.3$. The dotted, solid, and dashed lines correspond to $\omega_{de} = -1.2, -1.0$ and $-0.9$, respectively. As $\omega_{de}$ decreases, $\delta_{g}$ maintains the linear growth factor proportional to $a$ for a longer time. This is easy to understand. If there is the same amount of the matter at the present epoch for the different models, then there will be more matter component in the past for the smaller value of $\omega_{de}$. Thus, dark energy model with the smaller value of $\omega_{de}$ maintains the longer linear growth behavior.

We are also able to obtain the exact value of the growth index $f(a) = \fr{d \ln \delta_{g}}{d \ln a}$ in any epoch. We also get the exact value of the growth index parameter $\gamma(a) = \fr{ \ln f}{\ln \Omega_{m}(a)}$. We investigate the behaviors of $f(a)$ and $\gamma(a)$ based on this exact analytic growth factor for the different dark energy models as shown in Fig. \ref{fig2}.

\begin{center}
\begin{figure}
\vspace{1.5cm}
\centerline{
\psfig{file=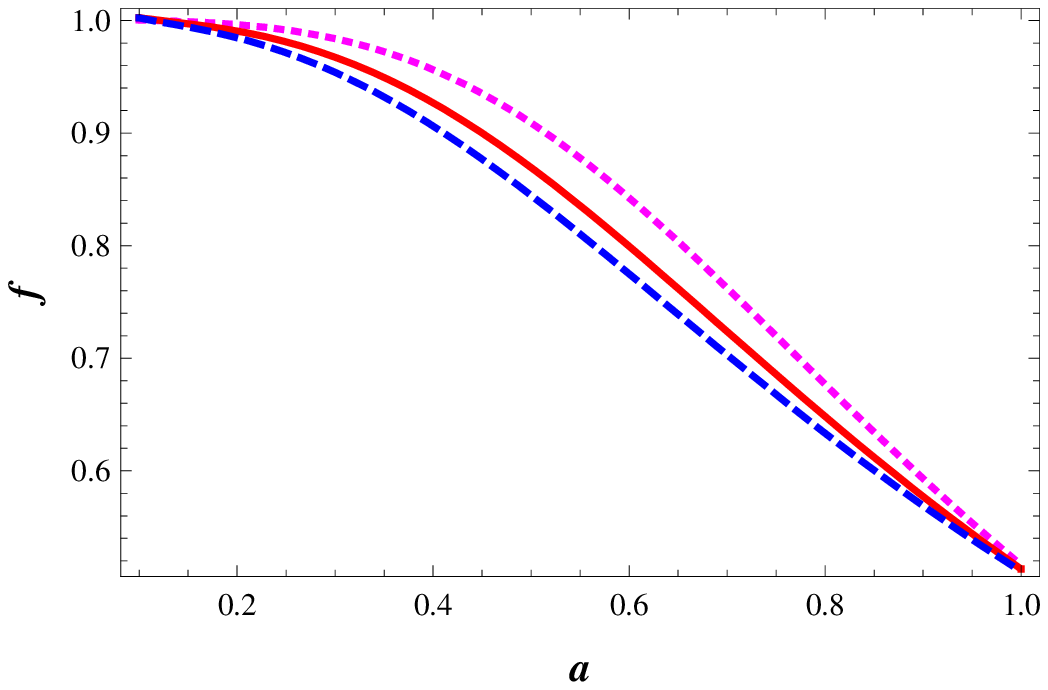, width=6cm} \psfig{file=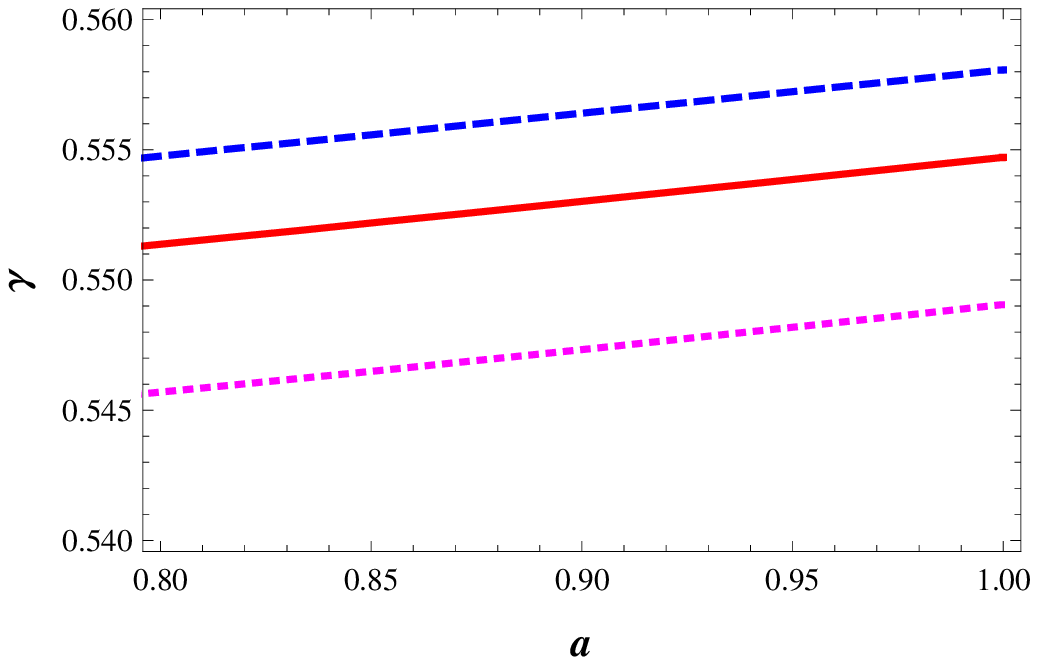, width=6cm} }
\vspace{-0.2cm}
\caption{ a) Evolution of $f(a)$ when $\Omega_{m}^{0} = 0.3$.  b) $\gamma(a)$ evolution with the same $\Omega_{m}^{0}$. Dashed, solid, and dotted lines correspond to $\omega_{de} = -0.9, -1.0$, and $-1.2$, respectively. } \label{fig3}
\end{figure}
\end{center}

In Fig. \ref{fig3}a, we show the cosmological evolution of $f(a)$ for the different dark energy models when $\Omega_{m}^{0} = 0.3$. Dashed, solid, and dotted lines correspond to $\omega_{de} = -0.9, -1.0$ and $-1.2$ dark energy models, respectively. As we show in Fig. \ref{fig3}a, the smaller value of $\omega_{de}$ gives the larger value of the present growth index $f(a=1)$. The present values of the growth index are $0.511, 0.513$, and $0.516$ for corresponding models, respectively. Thus we may not see any differences between different models from the present growth index. If we investigate the $f$ values at $z \simeq 0.15$ ({\it i.e.} $a \simeq 0.87$), then the growth index values will be $f(z = 0.15) = 0.588, 0.598$, and $0.618$. The current 2dF observation value is $ 0.36 \leq f(z=0.15) \leq 0.66$ \cite{Verde,Hawkins}.
In Fig. \ref{fig3}a, the behavior of $f(a)$ shows stronger model dependence around $0.25 \leq z \leq 1.5$. Thus, future observations in this redshift range should give us better observational constraints on the growth index.  In Fig. \ref{fig3}b, we show the evolution of the growth index parameter $\gamma(a)$ for corresponding models with the same notation as Fig. \ref{fig3}a. The present values $\gamma(a=1) = 0.5581, 0.5547$, and $0.5491$ for $\omega_{de} = -0.9, -1.0$, and $-1.2$, respectively. There is only $2 \%$ deviation from the different models. We are able to obtain the $\gamma$ values at $z = 0.15$ (equally $a \simeq 0.87$). We get $\gamma(z = 0.15) = 0.5559, 0.5525$, and $0.5468$ for corresponding models. This shows not much changing in $\gamma$ between the models. However, $\gamma$ values at high $z$ show very strong $\Omega_{m}^{0}$ dependence and vary dramatically \cite{SK}. Thus, we had better to limit $z < 2 \sim 3$ for the investigation of $\gamma$. We summarize the results in Table \ref{table1}. We also indicate the coefficients of the growth factor in this table, which are obtained from the growth factor initial conditions Eq. (\ref{ini}). In the last column, we compare the present growth index parameter  $\gamma_{WS}^{0}$ obtained from Ref. \cite{WS}. Those values are quite close to the ones obtained from the exact analytic solution. Thus, the approximation of $\gamma$ given in Ref. \cite{WS} is a good one as long as one probes $\gamma$ near the present.
\begin{center}
    \begin{table}
    \begin{tabular}{ | c | c | c | c | c | c | c | c|}
    \hline
    $\omega_{de}$ & $f_{(z=0)}$ & $\gamma_{(z=0)}$ & $f_{(z=0.15)}$ & $\gamma_{(z=0.15)}$ & $c_{1}$ & $c_{2}$  & $\gamma_{WS}^{0}$ \\ \hline
    -0.9 & 0.511 & 0.5581 & 0.588 & 0.5559  & -1.12344 & 1.08851 & 0.556 \\ \hline
    -1.0 & 0.513 & 0.5547 & 0.598 & 0.5525 & -0.943314 & 1.08464 & 0.553 \\ \hline
    -1.2 & 0.516 & 0.5491 & 0.618 & 0.5468  & -0.716894 & 1.07822 & 0.548 \\ \hline
    \end{tabular}
    \caption{$\omega_{de}$ is the eos of the dark energy. $f_{(z=0)}$ and $f_{(z=0.15)}$ are the values of the growth index at the present and $z = 0.15$, respectively. $\gamma_{(z=0)}$ and $\gamma_{(z=0.15)}$ are growth index parameters at the corresponding epoches. $c_{1}$ and $c_{2}$ are the coefficients of the growth factor $\delta_{g}$ obtained from the initial conditions in Eq. (\ref{ini}). $\gamma_{WS}^{0}$ are the values of the growth index parameters obtained from Ref. \cite{WS}. }
    \label{table1}
    \end{table}
\end{center}

In addition to this, if we naively take the surface values of the $1$-$\sigma$ result of the 2dF measurement without taking into account the uncertainties in the relevant cosmological parameters, then we may be able to rule out many dark energy models by using this exact analytic solution. For example, if $\omega_{de} = -3$ like in some phantom models, we will obtain $f(z=0.15) = 0.766$ for $\Omega_{m}^{0} = 0.3$, which is way too large compared to the 2dF observational value even when we consider the error in the data coming from the selection effects. Thus, dark energy models with too small values of $\omega_{de}$ should be ruled out if we want to keep the concordance model. However, there are still large uncertainties in the 2dF measurement and we need to wait for more accurate measurements to conclude this. Also some models with time varying $\omega_{de}$ might have better chance to survive.

It has been misunderstood that the sub-horizon scale growth factor for the general constant $\omega_{de}$ is known \cite{Silveira}. However, the solution in the given article is not correct because it claims that the growth and decaying solution are separable for the general $\omega_{de}$. It is well known that the separation of two modes is possible only when $\omega_{de} = $ $-\fr{1}{3}$ or $-1$ \cite{Dodelson}. The detail discussion about this is out of the main stream of letter \cite{SK3}.

Even though the results of this letter is limited for the constant equation of state of dark energy, we are able to apply these solutions to the time-varying $\omega_{de}$ by interpolating between models with constant $\omega_{de}$ \cite{SK3}. Also it is well known that the time-dependence of $\omega_{de}$ is extremely difficult to discern because the dark energy is dynamically unimportant at the redshifts where $\omega_{de}$ departs from its low $z$ value. In addition, for the substantial changes in $\omega_{de}$ at low redshift, there is always a constant $\omega_{de}$ that produces very similar evolution of all of the observables simultaneously \cite{Kujat,Maor2}. Also this analytic solution can provide useful templates to study the structure growth in dark energy models with time varying equation of state.

In our previous work, we also obtained the exact analytic solution of the growth factor $\delta_{g}^{L}$ for the cosmological constant case \cite{SK}. Even though, $\delta_{g}^{L}$ seems to be quite different from the solution $\delta_{g}$ in this letter, both are indeed same solutions which show the same physical behaviors. The details of comparison of them are irrelevant to the results of this letter and we do not show any detail in the present consideration \cite{SK3,SL}.


\begin{thebibliography}{99}

\bibitem{SK} S.~Lee and K.-W.~Ng, [astro-ph/0905.1522]

\bibitem{Bonnor} W.~B.~Bonnor, Mon.\ Not.\ R.\ Astron.\ Soc. {\bf 117}, 104 (1957).

\bibitem{Morse} P.~M.~Morse and H.~Feshbach, {\it Methods of Theoretical Physics, Part I} (McGraw-Hill Science, New York, 1953).

\bibitem{Verde} L.~Verde {\it et al.}, Mon.\ Not.\ R.\ Astron.\ Soc. {\bf 335}, 432 (2002) [astro-ph/0112161].

\bibitem{Hawkins} E.~Hawkins {\it et al.}, Mon.\ Not.\ R.\ Astron.\ Soc. {\bf 346}, 78 (2003) [astro-ph/0212375].

\bibitem{WS} L.~Wang and P.~J.~Steinhardt, Astrophys.\ J. {\bf 508}, 483 (1998) [astro-ph/9804015].

\bibitem{Silveira} V.~Silveira and I.~Waga, Phys.\ Rev.\ D {\bf 50}, 4890 (1994).

\bibitem{Dodelson} S.~Dodelson, {\it Modern Cosmology} (Academic Press, San Diego, 2002) Erratum-ibid.

\bibitem{SK3} S.~Lee and K.-W.~Ng, [arXiv:0907.2108].

\bibitem{Kujat} J.~Kujat, A.~M.~Linn, R.~J.Scherrer, and D.~H.~Weinberg, Astrophys.\ J.\ {\bf 572}, 1 (2002) [arXiv:astro-ph/0112221].

\bibitem{Maor2} I.~Maor, R.~Brustein, J.~McMahon, and P.~J.~Steinhardt, Phys.\ Rev.\ D {\bf 65}, 123003 (2002) [arXiv:astro-ph/0112526].

\bibitem{SL} S.~Lee, [arXiv:0905.4734].



\end{thebibliography}
\end{document}